\documentclass{article}
\usepackage[utf8]{inputenc}
\usepackage[english]{babel}
\usepackage{graphicx}
\usepackage{hyperref} 
\usepackage{geometry}
\usepackage{amsmath}
\usepackage{amssymb}
\usepackage{amsfonts}
\usepackage{multicol}
\usepackage{float}
\usepackage{caption}
\usepackage{subcaption}
\usepackage{cite}
\usepackage{amsthm}
\usepackage[utf8]{inputenc}
\usepackage{xcolor}
\def\@fmsl@sh#1#2#3{\m@th\ooalign{$\hfil#1\mkern#2/\hfil$\crcr$#1#3$}}
 \def\eq#1\en{\begin{equation}#1\end{equation}}
\def\s[#1,#2]{[#1\stackrel{\star}{,}#2]}
\def\sx[#1,#2]{[#1\stackrel{\star_{x}}{,}#2]}

\newcommand{\nc}{\newcommand}
\nc{\beq}{\begin{equation}}
\nc{\eeq}{\end{equation}}
\nc{\beqa}{\begin{eqnarray}}
\nc{\eeqa}{\end{eqnarray}}

\def\bc{\begin{center}}
\def\ec{\end{center}}

\def\gsim{\mathrel{\mathpalette\atversim>}}

\def\bc{\begin{center}}
\def\ec{\end{center}}

\def\gsim{\mathrel{\rlap{\lower4pt\hbox{\hskip1pt$\sim$}}

    \raise1pt\hbox{$>$}}}       %greater than or approx. symbol

\def\gsim{\mathrel{\rlap{\lower4pt\hbox{\hskip1pt$\sim$}}
    \raise1pt\hbox{$>$}}}       %greater than or approx. symbol

\begin{document}
\makeatletter
\def\fmslash{\@ifnextchar[{\fmsl@sh}{\fmsl@sh[0mu]}}
\def\fmsl@sh[#1]#2{%
  \mathchoice
    {\@fmsl@sh\displaystyle{#1}{#2}}%
    {\@fmsl@sh\textstyle{#1}{#2}}%
    {\@fmsl@sh\scriptstyle{#1}{#2}}%
    {\@fmsl@sh\scriptscriptstyle{#1}{#2}}}
\def\@fmsl@sh#1#2#3{\m@th\ooalign{$\hfil#1\mkern#2/\hfil$\crcr$#1#3$}}
\makeatother
%\baselineskip 24pt

%%%%%%%%%%%%%%%%%%%%%%%%%%%%%%%%%%%%%%%%%%%%%%%%%%%%%%%%%%%%%%%%%
%%%
%%%                      TITLE PAGE
%%%
%%%%%%%%%%%%%%%%%%%%%%%%%%%%%%%%%%%%%%%%%%%%%%%%%%%%%%%%%%%%%%%%%
\thispagestyle{empty}
\begin{titlepage}
\boldmath
\begin{center}
  \Large {\bf  Black Holes in Quantum Gravity}
    \end{center}
\unboldmath
\vspace{0.2cm}
\begin{center}
{{\large Xavier~Calmet}\footnote{E-mail: x.calmet@sussex.ac.uk}
{\large and}  {\large  Folkert~Kuipers}\footnote{E-mail: f.kuipers@sussex.ac.uk}}
 \end{center}
\begin{center}
{\sl Department of Physics and Astronomy,
University of Sussex, Brighton, BN1 9QH, United Kingdom
}\\
\end{center}
\vspace{5cm}
\begin{abstract}
\noindent
We review some recent results obtained for black holes using effective field theory methods applied to quantum gravity, in particular the unique effective action. Black holes are complex thermodynamical objects that not only have a temperature but also have a pressure. Furthermore, they have quantum hair which provides a solution to the black hole information paradox.
\end{abstract}
\vspace{5cm}
\end{titlepage}

%\pacs{}

%%%%%%%%%%%%%%%%%%%%%%%%%%%%%%%%%%%%%%%%%%%%%%%%%%%%%%%%%%%%%%%%
%%%
%%%                     INTRODUCTION
%%%
%%%%%%%%%%%%%%%%%%%%%%%%%%%%%%%%%%%%%%%%%%%%%%%%%%%%%%%%%%%%%%%%

\newpage
Black holes are fascinating objects which were posited long before the advent of general relativity. Indeed Michell in 1767 and Laplace 1799 had suggested independently that an astrophysical body could be so dense that the escape velocity of light would be such that it would not be able to leave the surface of the object which would thus be a black star.  Black holes were rediscovered in the twentieth century as classical solution of Einstein's equations of general relativity by Schwarzschild \cite{Schwarzschild:1916uq}. These static vacuum solutions have at least one horizon and a singularity. At a singularity, some of the curvature invariants become infinite. This can often be associated to the notion of geodesic incompleteness and signals a breakdown of the predictivity of general relativity. Two black hole solutions are particularly important: the Schwarzschild solution which represents a black hole of mass $M$ and the Kerr solution which represents a black hole of mass $M$ and angular momentum $J$.  The Schwarzschild metric has been studied extensively mainly because it is the simplest case possible of a black hole, but also because it reveals already many of the important features of the solutions. The Kerr metric \cite{Kerr:1963ud} is particularly important because it describes astrophysical black holes which are rotating objects. Charged black hole solutions, described by the Kerr-Newman solutions, also exist, but they may not be relevant to nature as black holes are thought to be electrically neutral. Black holes of a large range of potential masses have now been observed: from supermassive black holes nested at the center of galaxies, to stellar black holes. The existence of black holes in our universe has been further confirmed via the observation of black hole mergers and the production of gravitational waves produced during such mergers.

In this short paper, we introduce some key historical results on black holes before reviewing some recent results. We start with black holes in general relativity and introduce some key mathematical results on the properties of black holes in general relativity. We then extend our considerations to include quantum field theoretical results for black holes in a curved space-time which is still treated classically before coming to recent results to black holes in quantum gravity using an effective field theory approach.

\begin{itemize}
\item[A)] Black holes in general relativity 
\begin{itemize}

\item[1)] Birkhoff theorem \cite{Birkhoff} states that any spherically symmetric solution of the vacuum field equations must be static and asymptotically flat. This theorem implies that the exterior solution must be given by the Schwarzschild metric. 

\item[2)] No-hair theorem \cite{Israel:1967wq,Israel:1967za,Carter:1971zc} states that all black hole solutions of the Einstein-Maxwell equations of gravitation and electromagnetism in general relativity can be completely characterized by only three externally observable classical parameters: mass, electric charge, and angular momentum.

\item[3)] Penrose singularity theorem \cite{Penrose:1964wq} proves geodesic incompleteness of space-times that contain a trapped region.  As a result, this theorem implies that a singularity forms during the formation of black hole via gravitational collapse.
\end{itemize}

\item[B)] Black holes in general relativity and quantum mechanics
\begin{itemize}
\item[1)]  Hawking has shown that while black holes in general relativity are black, at the quantum mechanical level they emit a radiation similar to that of a black body. This implies that they have a temperature given by \cite{Hawking:1975vcx}
\begin{eqnarray}  \nonumber
T=\frac{1}{8 \pi G_N M}.
\end{eqnarray}
 \item[2)] Bekenstein and Hawking realized that black holes are complex thermodynamical systems that have an entropy \cite{Bekenstein:1973ur,Bekenstein:1972tm,Hawking1976}
\begin{eqnarray} \nonumber
S_{BH}  =\frac{A}{4 G_N},
\end{eqnarray}
where the black hole area $A=16 \pi (G_N M)^2$ ($G_N$ is Newton's constant and $M$ is the black hole mass).
\item[3)] Black hole information paradox: Hawking's black hole evaporation  \cite{Hawking1976}, see e.g. for reviews \cite{Mathur1,Mathur2,Raju:2020smc} for recent reviews, seems to imply that black holes cause pure states to evolve into mixed states. This would imply a non-unitarity evolution of the S-matrix and lead to a disaster for quantum mechanics.
\end{itemize}

\item[C)]{Black holes in quantum gravity}

It has long been speculated that the theorems A1, A2 and A3  may not hold in a theory of quantum gravity, but without a concrete theory of quantum gravity not much progress has been made along these lines. Recently, these questions have been revisited using techniques based on the concept of effective field theory. While we do not yet have a widely accepted theory of quantum gravity, model independent calculations in quantum gravity are nevertheless possible, if we assume that whatever the ultra-violet theory of quantum gravity might be, it has general relativity as its low energy limit. In particular, the unique effective action \cite{Weinberg:1980gg, Barvinsky:1984jd,Barvinsky:1985an,Barvinsky:1987uw,Barvinsky:1990up,Buchbinder:1992rb,Donoghue:1994dn} provides a gauge invariant and unique low energy theory of quantum gravity for any ultra-violet complete theory of quantum gravity which has general relativity as a low energy limit. At second order in curvature, one has
\begin{eqnarray}\label{EFTaction} \nonumber
S_{\text{EFT}} &=& \int \sqrt{|g|}d^4x  \left( \frac{R}{16\pi G_N} + c_1(\mu) R^2 + c_2(\mu) R_{\mu\nu} R^{\mu\nu} + c_3(\mu) R_{\mu\nu\rho\sigma} R^{\mu\nu\rho\sigma} + \mathcal{L}_m \right) \ ,
\end{eqnarray}
for the local part of the action and the nonlocal part is given by
\begin{eqnarray}\label{nonlocalaction} \nonumber
	\Gamma_{\text{NL}}^{\scriptstyle{(2)}}  &=& - \int  \sqrt{|g|}d^4x \left[ b_1 R \ln\left(\frac{\Box}{\mu^2}\right)R + b_2 R_{\mu\nu} \ln\left(\frac{\Box}{\mu^2}\right) R^{\mu\nu} + b_3 R_{\mu\nu\alpha\beta} \ln\left(\frac{\Box}{\mu^2}\right)R^{\mu\nu\alpha\beta} \right],
	\end{eqnarray}
	where $\Box := g^{\mu\nu} \nabla_\mu \nabla_\nu$ and $\mu$ is a renormalization scale. While the Wilson coefficients $c_i$ of the local operators are not calculable without assuming an ultra-violet complete theory of quantum gravity, those of the non-local operators can be calculated in a model independent way. Indeed, the coefficients $b_i$ depend on the number of fields with spin 0, 1/2, 1 and 2 included in the model. Within this framework, one can investigate quantum gravitational corrections to classical solutions of general relativity by calculating corrections to these solutions using perturbation theory. Using this framework, we have revisited classical black hole results and found:
\begin{itemize}
\item[1)] Penrose's singularity theorem does not hold in quantum gravity \cite{Kuipers:2019qby}. Indeed, to second order in curvature, the effective quantum gravitational action contains two new degrees of freedom which have important implications for the necessary conditions of this theorem: a massive spin-2 field and a massive spin-0 field. Using an explicit mapping of this theory from the Jordan frame to the Einstein frame, it can be shown that the massive spin-2 field violates the null energy condition, while the massive spin-0 field satisfies the null energy condition, but violates the strong energy condition. Due to the violation of the null-energy condition, Penrose's singularity theorem is no longer applicable, indicating that singularities can be avoided, if the leading quantum corrections are taken into account.

\item[2)] Birkhoff's theorem does not hold in quantum gravity: It can be shown explicitly \cite{Calmet:2017qqa,Calmet:2021lny}  that there are quantum gravitational corrections to third order in curvature to the classical Schwarzschild solution. Moreover, new branches of solutions exist, if the leading quantum gravitational corrections are incorporated \cite{Lu:2015cqa,Lu:2015psa}.

\item[3)] Black hole thermodynamics: we can calculate quantum gravitational corrections to the entropy of black holes using effective field theory techniques. The second order correction in curvature to the entropy is given by \cite{Calmet:2021lny}. 
\begin{eqnarray} \label{2ndentropy} \nonumber
S_{Wald}^{(2)}  &=&\frac{A}{4 G_N} +  64 \pi^2 c_3(\mu) +64 \pi^2 b_3 \left( \log \left (4 G^2_N M^2 \mu^2\right) -2 +2 \gamma_E \right),
\end{eqnarray}
where $A$ is the area of the black hole obtained by Bekenstein and Hawking. Our result implies that black holes not only have a temperature, but also a pressure given by 
\begin{eqnarray} \label{pressure} \nonumber
P=-b_3 \frac{1}{2 G_N^4 M^4},
\end{eqnarray}
which can be negative as $b_3$ is positive for spin 0, 1/2 and 2 fields or positive as $b_3$ is negative for spin 1 fields. Indeed, one finds  $b_{3,0}=2/(11520 \pi^2)$ \cite{Deser:1974cz},  $b_{3,1/2}=7/(11520 \pi^2)$ \cite{Deser:1974cz}, $b_{3,1}=-26/(11520 \pi^2)$ \cite{Deser:1974cz} and $b_{3,2}=424/(11520 \pi^2)$ \cite{Barvinsky:1984jd}. 

\item[4)] The no-hair theorem does not hold in quantum gravity: it can be shown explicitly \cite{Calmet:2019eof,Calmet:2021stu} that the quantum state of the graviton field outside a compact object carries information about the internal state of the object.  In general relativity, an external observer cannot differentiate a star with radius $R_s$ and mass $M$ from the star with two different components but same external radius and same total mass $M$. It can be shown using effective field theory methods \cite{Calmet:2021stu} that the quantum gravitational corrections are different for the two matter distributions and there is thus a memory effect when the stars collapse to black holes: black holes have quantum hair.

\item[5)] The quantum hair leads to the realization that black hole evaporation is unitary \cite{Calmet:2021cip}: the quantum state of the graviton field outside a black hole horizon carries information about the internal state of the hole. This allows unitary evaporation: the final radiation state is a complex superposition which depends linearly on the initial black hole state. Under time reversal, the radiation state evolves back to the original black hole quantum state.
\end{itemize}

\end{itemize}

In conclusion, black holes are fascinating objects firstly discovered purely theoretically. Their history dates back to the eighteenth century  and they were rediscovered in general relativity in the twentieth century. By now, there is ample observational evidence that black holes exist in our universe. Black holes are particularly interesting as a theoretical laboratory to try to understand quantum mechanical effects in general relativity. We have seen how recent progress in quantum gravity enables us to revisit many classical results in black hole physics. In particular, we can now show that black hole evaporation is a unitary process.

\bigskip

{\it Acknowledgments:}
The work of X.C. is supported in part  by the Science and Technology Facilities Council (grants numbers ST/T00102X/1 and ST/T006048/1).

%%%%%%%%%%%%%%%%%%%%%%%%%%%%%%%%%%%%%%%%%%%%%%%%%%%%%%%%%%%%%%%%%
%%%
%%%                     BIBLIOGRAPHY
%%%
%%%%%%%%%%%%%%%%%%%%%%%%%%%%%%%%%%%%%%%%%%%%%%%%%%%%%%%%%%%%%%%%%

\end{document}